\documentclass[10pt,conference]{IEEEtran}
\IEEEoverridecommandlockouts

\usepackage[utf8]{inputenc} 
\usepackage[T1]{fontenc}
\usepackage{url}
\usepackage{ifthen}
\usepackage{cite}
\usepackage{algorithmicx}
\usepackage{algorithm}
\usepackage{graphicx}
\usepackage{textcomp}
\usepackage{xcolor}
\usepackage{multirow}
\usepackage{multicol}
\usepackage{caption}
\usepackage{subcaption}
\usepackage{amsthm}
\usepackage{amssymb}
\usepackage{svg}
\usepackage{stfloats}
\usepackage[noend]{algpseudocode}

\usepackage[cmex10]{amsmath} 

\theoremstyle{definition}
\newtheorem{theorem}{Theorem}
\newtheorem{lemma}{Lemma}

\allowdisplaybreaks

\DeclareMathSizes{10}{9}{6.5}{5} 
\begin{document}
    \title{QML-IB: Quantized Collaborative Intelligence between Multiple Devices and the Mobile Network} 




\author{\IEEEauthorblockN{
Jingchen Peng\IEEEauthorrefmark{1}, 
Boxiang Ren\IEEEauthorrefmark{1}, 
Lu Yang\IEEEauthorrefmark{2}$^\ddag$, 
Chenghui Peng\IEEEauthorrefmark{2},
Panpan Niu\IEEEauthorrefmark{1},
and Hao Wu\IEEEauthorrefmark{1}}
\IEEEauthorblockA{\IEEEauthorrefmark{1}Department of Mathematical Sciences, Tsinghua University, Beijing, China}
\IEEEauthorblockA{\IEEEauthorrefmark{2}Wireless Technology Lab, Central Research Institute, 2012 Labs, Huawei Tech. Co. Ltd., China}
  \IEEEauthorblockA{Email: yanglu87@huawei.com}
\thanks{The first two authors contributed equally to this work and $\ddag$ marked the corresponding author. }
 }



\maketitle


\begin{abstract}
The integration of artificial intelligence (AI) and mobile networks is regarded as one of the most important scenarios for 6G. In 6G, a major objective is to realize the efficient transmission of task-relevant data. Then a key problem arises, how to design collaborative AI models for the device side and the network side, so that the transmitted data between the device and the network is efficient enough, which means the transmission overhead is low but the AI task result is accurate. In this paper, we propose the multi-link information bottleneck (ML-IB) scheme for such collaborative models design. We formulate our problem based on a novel performance metric, which can evaluate both task accuracy and transmission overhead. Then we introduce a quantizer that is adjustable in the quantization bit depth, amplitudes, and breakpoints. Given the infeasibility of calculating our proposed metric on high-dimensional data, we establish a variational upper bound for this metric. However, due to the incorporation of quantization, the closed form of the variational upper bound remains uncomputable. Hence, we employ the Log-Sum Inequality to derive an approximation and provide a theoretical guarantee. Based on this, we devise the quantized multi-link information bottleneck (QML-IB) algorithm for collaborative AI models generation. Finally, numerical experiments demonstrate the superior performance of our QML-IB algorithm compared to the state-of-the-art algorithm.




\end{abstract}

\section{Introduction}
As artificial intelligence (AI) undergoes a resurgence, its applications are expanding across various domains, including virtual/augmented reality(VR/AR) \cite{hou2018predictive}, \cite{anthes2016state}, speech recognition \cite{gaikwad2010review}, natural language processing \cite{chowdhary2020natural}, etc. The integration of AI and communication has been proposed as one of the most important usage scenarios of the sixth generation (6G) mobile networks by the International Telecommunication Union (ITU) \cite{recommendation2023framework}. In other words, future mobile networks should support AI-enabled functions natively, providing a seamless integration of sensing, communication, computation, and intelligence \cite{letaief2021edge}. \par
In traditional mobile networks, how to realize fast and complete data transmission is one of the core problems in network design. While in 6G, there exists a fact that different AI tasks may need different kinds of data. In addition, as the data increases explosively, whether all these data should be transmitted completely as before is controvertible. A hot viewpoint recently is that only the task-related data should be transmitted in future mobile networks \cite{strinati20216g,zhu2020toward,jankowski2020wireless}. Then, an important scenario appears, that only the AI task-related data in the device should be extracted, and transmitted over the wireless links to the network side. In this scenario, lots of important problems need to be solved, including how to design the device-network collaborative AI models \cite{kang2017neurosurgeon,zhou2019edge,shao2021branchy}, how to design the performance metric to evaluate the quality of transmitted data, etc. \par
In practical applications such as robotics \cite{zou2019collaborative},\cite{coppola2020survey}, security camera systems \cite{liu2016deep},\cite{tang2019cityflow}, and smart drone swarms \cite{unlu2019deep}, collaborative AI task execution across multiple devices is an important scenario. Current research, however, mainly focuses on single device-network collaboration, with scant attention to the collaboration between the network side and multiple mobile devices. 
One of the core reasons is the lack of a proper performance metric for the latter scenario, which should effectively evaluate both the AI task performance as well as the communication costs across multiple device-network links. The study \cite{shao2022task} proposed an algorithm for multi-device and network cooperation, based on the distributed information bottleneck (DIB) theory \cite{aguerri2019distributed}. 
However, this algorithm does not consider the impact of channel noise, and their AI models in the device side and network side cannot be trained collaboratively. More specifically, both of their AI models and their quantization scheme at the device side are fixed, and cannot adapt to the dynamic environment. In addition, due to the intrinsic data distribution among sensors and nodes in the network, the DIB was extended to networks modeled by directed graphs to enable distributed learning\cite{moldoveanu2021network},\cite{moldoveanu2023network}. \par

In this paper, we develop a multi-link information bottleneck (ML-IB) scheme to design the device-network collaborative AI models across multiple wireless links as illustrated in Fig.~\ref{fig: system model}. Here, we first propose a performance metric $\mathcal{C}_{\text{ML-IB}}$, inspired by the information bottleneck (IB) theory \cite{tishby2000information}, and formulate our objective function based on it. This metric can evaluate the AI task accuracy, together with transmission overhead, which can provide a comprehensive measure of how well a collaborative intelligent system between multiple devices and the network side\footnote{The network side can refer to the base station (BS), the edge server, etc. In this paper, we will mainly use BS to represent the network side for brevity.} performs. Next, we devise a simple yet efficient quantization scheme to ensure the compatibility between our models and digital communication systems. Third, in order to design our device-network collaborative AI models based on the aforementioned problem formulation, $\mathcal{C}_{\text{ML-IB}}$ should be computable. To tackle this problem, we utilize the standard variational methods and deduce a variational upper bound. 
Nevertheless, the integration of the quantization approach continues to pose computational difficulties in the estimation of the variational upper bound. We derive a novel approximation for the variational upper bound, denoted as $\mathcal{C}_{\text{QML-IB}}$. 
We also provide an error analysis of this approximation for theoretical guarantee. Then based on $\mathcal{C}_{\text{QML-IB}}$, we propose the corresponding QML-IB algorithm, and generate the collaborative AI models for both multiple devices and the BS. Finally, our numerical experiments demonstrate the effectiveness of our algorithm, which outperforms the state-of-the-art method in terms of the accuracy of AI tasks.
\section{System Model and Problem Formulation}
\begin{figure}[htbp]
\raggedright
\includegraphics[width=\linewidth]{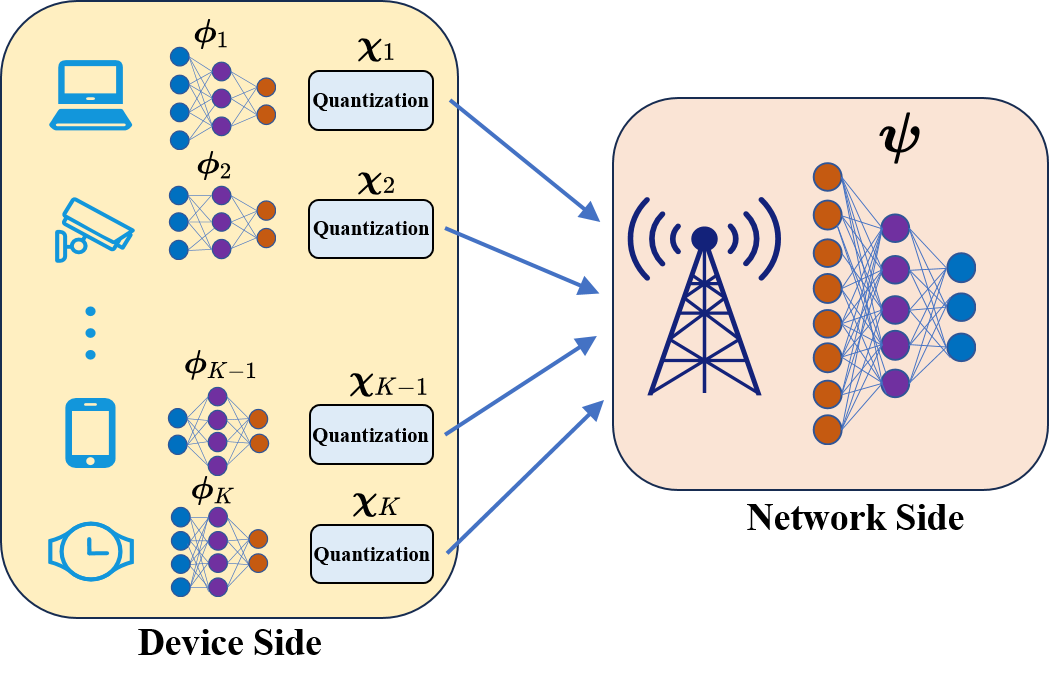}
\caption{System model of collaborative intelligence between multiple devices and the network.}
\label{fig: system model}
\end{figure}

In this section, we will introduce the system model of the device-network collaborative intelligence system containing multiple wireless 
links, with a quantization unit for practical considerations, 
as shown in Fig.~\ref{fig: system model}. Based on it, we design a performance metric $\mathcal{C}_{\text{ML-IB}}$ and formulate the optimization problem. Using the standard variational techniques in \cite{kingma2013auto} and \cite{alemi2016deep}, we address the mutual information terms in $\mathcal{C}_{\text{ML-IB}}$.



\subsection{System Model and Data Transmission Chains}\label{subsec: system model}
The input data of $K$ different mobile devices $(\boldsymbol{x}_1,\dots,\boldsymbol{x}_K)$ and the target variable $\boldsymbol{y}$ (e.g., label) are deemed as different realizations of the random variables $(X_1,\dots, X_K, Y)$ with joint distribution $p(\boldsymbol{x}_1,\dots,\boldsymbol{x}_K,\boldsymbol{y})$. Given the dataset $\{\boldsymbol{x}^{(m)}_{1},\dots,\boldsymbol{x}^{(m)}_{K}\}_{m=1}^M$, the goal of the system is to infer the targets $\{\boldsymbol{y}^{(m)}\}_{m=1}^M$. 
Upon receiving the input data $\boldsymbol{x}_k$, the $k$-th device extracts features from $\boldsymbol{x}_k$ and encodes them as $\boldsymbol{z}_k$. 
We collectively denote the feature extraction and encoding modules as the probability encoder $p_{\boldsymbol{\phi}_k}(\boldsymbol{z}_k|\boldsymbol{x}_k)$, with the trainable DNN parameters $\boldsymbol{\phi}_k$ for the $k$-th device. To provide a clear probabilistic model for the devices, we employ the reparameterization trick \cite{kingma2015variational}. We then model the conditional distribution $p_{\boldsymbol{\phi}_k}(\boldsymbol{z}_k|\boldsymbol{x}_k)$ as a multivariate Gaussian distribution, that is, 
\begin{equation}\label{eq: conditional x_k to z_k}
p_{\boldsymbol{\phi}_k}(\boldsymbol{z}_k|\boldsymbol{x}_k) = \mathcal{N}(\boldsymbol{z}_k|\boldsymbol{\mu}_k,\boldsymbol{\Theta}_k).
\end{equation}
Here, the covariance matrix $\boldsymbol{\Theta}_k$ is a diagonal matrix $\mathrm{diag} \{\theta_{k,1}^2,\dots,\theta_{k,d}^2\}$.
The mean vector $\boldsymbol{\mu}_k = (\mu_{k,1},\dots,\mu_{k,d})$ and the diagonal vector $\boldsymbol{\theta}_k = (\theta_{k,1},\dots, \theta_{k,d})$ are derived from the DNN-based function $\boldsymbol{\mu}_k(\boldsymbol{x}_k,\boldsymbol{\phi}_k)$
and $\boldsymbol{\theta}_k(\boldsymbol{x}_k,\boldsymbol{\phi}_k)$, respectively. \par
Next, the encoded feature $\boldsymbol{z}_k$ is quantized to $\tilde{\boldsymbol{z}}_k$ using our quantization function, which will be specified later, with all trainable parameters denoted as $\boldsymbol{\chi}_k$. The quantized feature $\tilde{\boldsymbol{z}}_{k}$ is then transmitted to the BS by the $k$-th device through the wireless channel. 
Without loss of generality, we consider the additive white Gaussian noise channel. Thus, $\hat{\boldsymbol{z}}_k = \tilde{\boldsymbol{z}}_k + \boldsymbol{\epsilon}_k$, where the noise $\boldsymbol{\epsilon}_k \sim \mathcal{N}(\boldsymbol{0},\sigma_k^2 \boldsymbol{I})$ with its variance $\sigma_k^2$. The encoded feature $\boldsymbol{z}_k$, the quantized feature $\tilde{\boldsymbol{z}}_k$, and the received feature $\hat{\boldsymbol{z}}_k$ are instantiations of the random variables $Z_k$, $\tilde{Z}_k$, and $\hat{Z}_k$, respectively, allowing us to construct a Markov chain as
\begin{equation}\label{eq: Markov chain}
    Y \leftrightarrow X_k \leftrightarrow Z_k \leftrightarrow \tilde{Z}_k \leftrightarrow \hat{Z}_k.
\end{equation}
Here, the probability equation is given by $p(\hat{\boldsymbol{z}}_k,\tilde{\boldsymbol{z}}_k,\boldsymbol{z}_k|\boldsymbol{x}_k) =p(\hat{\boldsymbol{z}}_k|\tilde{\boldsymbol{z}}_k)p(\tilde{\boldsymbol{z}}_k|\boldsymbol{z}_k)p_{\boldsymbol{\phi}_k} (\boldsymbol{z}_k|\boldsymbol{x}_k)$. The network side performs inference tasks based on the received features $(\hat{\boldsymbol{z}}_1,\dots,\hat{\boldsymbol{z}}_K)$ from $K$ different devices, resulting in the inference output $\hat{\boldsymbol{y}}$. This output follows the 
conditional distribution $p_{\boldsymbol{\psi}}(\hat{\boldsymbol{y}}|\hat{\boldsymbol{z}}_{1:K})$, where $\boldsymbol{\psi}$ represents the parameters of the DNN employed at the network side.\par

\subsection{The ML-IB Scheme}
For the system shown in Fig.~\ref{fig: system model}, our objective is to design proper AI models for both the device side and the network side, which can ensure the accuracy of device-network collaborative AI tasks is high while the amount of data transmitted over wireless links is low. In the single device-network collaboration scenario,  conventional IB theory offers an effective approach to achieve this objective \cite{shao2021learning}. Based on IB, we develop a multi-link information bottleneck (ML-IB) scheme for the scenario of intelligent collaboration between multiple devices and the network side. Initially, a novel performance metric is proposed, defined as:
\begin{equation}\label{eq: ML-IB}
\mathcal{C}_{\text{ML-IB}}=-I(Y;\hat{Z}_{1:K}) + \sum_{k=1}^K \beta_k I(X_k;\hat{Z}_k).
\end{equation}\par
In this metric, the mutual information $I(Y;\hat{Z}_{1:K})$ reflects the information that received features of the network side $\hat{Z}_{1:K} = \{\hat{Z}_1, \dots, \hat{Z}_K\}$ holds about the inference target variable $Y$, which can represent the inference accuracy. \(I(X_k;\hat{Z}_k)\) assesses the retained information in \(\hat{Z}_k\) given $X_k$ using the minimum description length \cite{cover1999elements}, representing communication overhead. In addition, it can reflect the impact of encoding, quantization, and transmission processes on the input raw data. The tradeoff factor $\beta_k>0 $ can be dynamically adjusted based on the specific conditions of the $k$-th channel. The summation $\sum_{1}^K \beta_k I(X_k;\hat{Z}_k)$ can represent the communication overhead of all of the $K$ wireless links. As such, a smaller \(\mathcal{C}_{\text{ML-IB}}\) corresponds to a higher accuracy of the AI task and a lower communication overhead. The objective function can thus be formulated as
\begin{equation}\label{eq: problem}
    \min_{\{\boldsymbol{\phi}_k\}_{k=1}^K,\{\boldsymbol{\chi}_k\}_{k=1}^K,\boldsymbol{\psi}} \mathcal{C}_{\text{ML-IB}},
\end{equation}
where we aim to determine the optimal AI models for the multiple devices (i.e., $\{\boldsymbol{\phi}_k\}_{k=1}^K$), the network side (i.e., $\boldsymbol{\psi}$), and the optimal quantization framework (i.e., $\{\boldsymbol{\chi}_k\}_{k=1}^K$) to achieve minimum $\mathcal{C}_{\text{ML-IB}}$.\par

\subsection{Variational Upper Bound of $\mathcal{C}_{\text{ML-IB}}$}
To solve the problem \eqref{eq: problem}, in this subsection, we focus on analyzing the expression of $\mathcal{C}_{\text{ML-IB}}$. In order to calculate the mutual information terms in $\mathcal{C}_{\text{ML-IB}}$, we need the conditional distribution $p(\boldsymbol{y}|\hat{\boldsymbol{z}}_{1:K})$ for $I(Y;\hat{Z}_{1:K})$ and the marginal distribution $p(\hat{\boldsymbol{z}}_k)$ for $I(X_k;\hat{Z}_k)$.
However, the high dimensionality of the input data poses inherent computational challenges, making accurate computation of these distributions difficult. To overcome this, we resort to variational approximation \cite{kingma2013auto}, \cite{alemi2016deep}, which is a popular technique used to approximate intractable distributions based on some adjustable parameters (e.g., weights in DNNs). We introduce variational distributions $\{r_k(\hat{\boldsymbol{z}}_k)\}_{k=1}^K$ as approximations for $\{p(\hat{\boldsymbol{z}}_k)\}_{k=1}^K$, where each  $r_k(\hat{\boldsymbol{z}}_k)$ is modeled as a centered isotropic Gaussian distribution \cite{alemi2016deep}, represented by $\mathcal{N}(\hat{\boldsymbol{z}}_k|\boldsymbol{0},\boldsymbol{I})$. Since the inference variable $\hat {Y}$ and the target variable $Y$ share the same range of values, we use $p_{\boldsymbol{\psi}}(\boldsymbol{y}|\hat{\boldsymbol{z}}_{1:K})$
as the variational approximation for $p(\boldsymbol{y}|\hat{\boldsymbol{z}}_{1:K})$, with $\boldsymbol{\psi}$ denoting the DNN parameters on the network side.\par
According to the non-negativity of the KL divergence 
$D_{\text{KL}}(p(\boldsymbol{y}|\hat{\boldsymbol{z}}_{1:K}),p_{\boldsymbol{\psi}}(\boldsymbol{y}|\hat{\boldsymbol{z}}_{1:K})) \geq 0$, we derive a lower bound for $I(Y;\hat{Z}_{1:K})$ as :
\begin{equation}\label{eq: upbound1}
    I(Y;\hat{Z}_{1:K}) 
    \geq \int p(\hat{\boldsymbol{z}}_{1:K},\boldsymbol{y})\log \frac{p_{\boldsymbol{\psi}}(\boldsymbol{y}|\hat{\boldsymbol{z}}_{1:K})}{p(\boldsymbol{y})} \, d\hat{\boldsymbol{z}}_{1:K} \, d\boldsymbol{y}.
\end{equation}
Similarly, the non-negativity of \( D_{\text{KL}}(p(\hat{\boldsymbol{z}}_k),r_k(\hat{\boldsymbol{z}}_k)) \) allows us to deduce:
\begin{equation}\label{eq: upbound2}
    I(\hat{Z}_k;X_k) \leq \int p(\hat{\boldsymbol{z}}_k,\boldsymbol{x}_k)\log\frac{p_{\boldsymbol{\phi}_k}(\hat{\boldsymbol{z}}_k|\boldsymbol{x}_k)}{r_k(\hat{\boldsymbol{z}}_k)}d\hat{\boldsymbol{z}}_kd\boldsymbol{x}_k.
\end{equation}\par
By combining \eqref{eq: upbound1} and \eqref{eq: upbound2}, we can derive an upper bound of $\mathcal{C}_{\text{ML-IB}}$ ignoring a constant term $H(Y)$ as
\begin{equation}\label{eq: VML-IB}
\begin{aligned}
    \hat{\mathcal{C}}_{\text{ML-IB}} &= \mathbb{E}_{p(\boldsymbol{x}_{1:K},\boldsymbol{y})} \Big \{ \mathbb{E}_{p(\hat{\boldsymbol{z}}_{1:K}|\boldsymbol{x}_{1:K})}[-\log p_{\boldsymbol{\psi}}(\boldsymbol{y}|\hat{\boldsymbol{z}}_{1:K})] \\
\quad\! &\quad+\! \sum_{k=1}^K \beta_k D_{\text{KL}}(p_{\boldsymbol{\phi}_k}(\hat{\boldsymbol{z}}_k|\boldsymbol{x}_k)||r_k(\hat{\boldsymbol{z}}_k))  \Big \}.
\end{aligned}
\end{equation}

\section{
Computable ML-IB Model
}\label{sec: quantization}

The numerical calculation of the upper bound $\hat{\mathcal{C}}_{\text{ML-IB}}$ is still difficult, due to the intractable KL divergence term when the quantization units are considered. 

Following the idea in \cite{ye2021autoencoder}, we design the quantization unit. Our quantizer allows more flexible control of the quantization bit depth, breakpoint, and amplitude, which takes real-world deployment into account. To tackle the upper bound $\hat{\mathcal{C}}_{\text{ML-IB}}$, we derive a manageable approximation of $\hat{\mathcal{C}}_{\text{ML-IB}}$ by applying the Log-Sum Inequality, enabling the computability of our ML-IB scheme. Furthermore, we provide an error bound for the approximation to provide a theoretical guarantee.

\subsection{
Quantization Design
}\label{subsec: quantizion details}
We denote the encoded vector \( \boldsymbol{z}_k = (z_{k,1}, \dots, z_{k,d}) \in \mathbb{R}^d\). 
For the quantizer, we utilize a piecewise step function defined by:
\begin{equation}\label{eq: quantization}
\tilde{z}_{k,i} = \sum_{t=1}^T \frac{a_{k,t}^2}{\sum_{t=1}^T a_{k,t}^2} \text{sgn}(z_{k,i} - b_{k,t}),\quad i=1,\dots,d,
\end{equation}
where $T$ reflects the number of quantization bits, $a_{k,t}$ represents the amplitude and $b_{k,t}$ is the breakpoint. 
Then the quantizer transforms each dimension $z_{k,i}$ of the vector $\boldsymbol{z}_k$ into $\tilde{z}_{k,i}$, which can take only $T+1$ discrete values between $-1$ and $1$. For brevity, let us denote $c_{k,1} = -1,c_{k,2} = \frac{a_{k,1}^2-\sum_{t=2}^T a_{k,t}^2}{\sum_{t=1}^{T} a_{k,t}^2},\dots, c_{k,T+1}=1$. 
We specify the quantization function for $\tilde{z}_{k,i}$ as follows:
\begin{small}
\begin{equation}
	\tilde{z}_{k,t} = \begin{cases}
	c_{k,1}, &\text{if}~ z_{k,i} < b_{k,1},\\
	c_{k,2}, &\text{if}~ b_{k,1} \leq z_{k,i} < b_{k,2},\\
        &\cdots  \\
        c_{k,T+1} &\text{if}~  z_{k,i} \geq b_{k,T}.\\
		   \end{cases}
\end{equation}
\end{small}
\par
As introduced in Section $\ref{subsec: system model}$, each component ${Z}_{k,i}$ of the variable ${Z}_k$ is conditionally independent given $X_k$.
Consequently, the conditional cumulative distribution function of $Z_{k,i}$, conditioned on $X_k$ is obtained as
\begin{equation}
\begin{aligned}    
F_{Z_{k,i}|X_k}(c)&=P(Z_{k,i}<c|X_k)\\
&=\frac{1}{2} \left[1 + \text{erf}\left(\frac{c - \mu_{k,i}}{\theta_{k,i}\sqrt{2}}\right)\right].
\end{aligned}
\end{equation}
From above, we denote $P(\tilde{Z}_{k,i}=c|X_k) = P_{k,i}(c)$ and derive the conditional probability mass function of the quantized variable $\tilde{Z}_{k,i}$ given $X_k$, as described below:
\begin{small}
\begin{equation}
	\!\!P_{k,i}(c) \!=\! \begin{cases}
	F_{Z_{k,i}|X_k}(b_{k,1}), &\text{if}~ c = c_{k,1},\\
	F_{Z_{k,i}|X_k}(b_{k,2})-F_{z_{k,i}|\boldsymbol{x}_k}(b_{k,1}), &\text{if}~ c=c_{k,2},\\
        &\cdots  \\
        1-F_{Z_{k,i}|X_k}(b_{k,T}), &\text{if}~  c=c_{k,T+1}.\\
		   \end{cases}
\end{equation}
\end{small}
\par

Then, we consider a soft quantization function, which is defined by the following equation for each component $\tilde{z}_{k,i}$:
\begin{equation} \label{eq: soft-quantization}
    \tilde{z}_{k,i}=\sum_{t=1}^T \frac{a^2_{k,t}}{\sum_{t=1}^{T} a_{k,t}^2}\arctan(\gamma(z_{k,i}-b_{k,t})).
\end{equation}
In this formulation, the parameters $a_{k,t}$ and $b_{k,t}$ for $t=1$ to $T$, belonging to $\boldsymbol{\chi}_k$, are adaptable during training to reflect real-time conditions. This makes the quantization function effectively adapt to
various channels and different transmission devices.
\par
Given $\hat{\boldsymbol{z}}_k=\tilde{\boldsymbol{z}}_k+\boldsymbol{\epsilon}_k$ with $\boldsymbol{\epsilon}_k \sim \mathcal{N}(0,\sigma_k^2 I)$, each component $\hat{z}_{k,i}$ independently follows a conditionally independent Gaussian mixture conditional on $X_k$, for $i=1,\dots,d$. To be specific, the conditional probability density function for $\hat{z}_{k,i}$ is expressed as:
\begin{equation}\label{eq: conditional distribution}
p_{\boldsymbol{\phi}_k}(\hat{z}_{k,i}|\boldsymbol{x}_{k})=\sum_{t=1}^{T+1} P_{k,i}(c_{k,t})f_{\epsilon}(\hat{z}_{k,i}-c_{k,t}),
\end{equation}
where \( f_{\epsilon} \) denotes the density function of \( \epsilon_{k,i} \), and \( \epsilon_{k,i} = \hat{z}_{k,i}-c_{k,t} \) follows a normal distribution \( \mathcal{N}(0,\sigma_k^2) \).
\subsection{Approximation of Variational Upper bound}\label{subsec: quantization upperbound}
The computation of the variational bound $\hat{\mathcal{C}}_{\text{ML-IB}}$, involving the integral of logarithms of a Gaussian mixture as indicated by $p_{\boldsymbol{\phi}_k}(\hat{z}_{k,i}|\boldsymbol{x}_{k})$ in \eqref{eq: conditional distribution}, presents a notable challenge.
In this subsection, We produce an approximation of $\hat{\mathcal{C}}_{\text{ML-IB}}$ with a theoretical guarantee, making our ML-IB scheme computable.\par
To begin, we define
\begin{equation}\label{eq: upper KL}
\begin{aligned}
    &D^*_{\text{KL}}(p_{\boldsymbol{\phi}_k}(\hat{\boldsymbol{z}}_k|\boldsymbol{x}_k)||r_k(\hat{\boldsymbol{z}}_k))\\
    &\triangleq -H(\tilde{Z}_{k})+\sum_{i=1}^d \Big\{ \log(T+1)-\frac{1}{2}\log{\sigma_{k}^2}\\
    &\qquad\qquad \qquad  \qquad +\frac{1}{2}\sigma_{k}^2-\frac{1}{2}+\sum_{t=1}^{T+1} P_{k,i}(c_{k,t})(c_{k,t})^2 \Big\},\\
    \end{aligned}
\end{equation}
and 
\begin{equation}\label{eq: QVMDN-IB}
\begin{aligned}
\mathcal{C}_{\text{QML-IB}} 
&\triangleq\! \mathbb{E}_{p(\boldsymbol{x}_{1:K},\boldsymbol{y})} \Big \{ \mathbb{E}_{p(\hat{\boldsymbol{z}}_{1:K}|\boldsymbol{x}_{1:K})}[-\log p_{\boldsymbol{\psi}}(\boldsymbol{y}|\hat{\boldsymbol{z}}_{1:K})] \\
&\quad\! +\! \sum_{k=1}^K \beta_k D^*_{\text{KL}}(p_{\boldsymbol{\phi}_k}(\hat{\boldsymbol{z}}_k|\boldsymbol{x}_k)||r_k(\hat{\boldsymbol{z}}_k))  \Big \}.\\
\end{aligned}
\end{equation}  

\begin{lemma}\label{lemma: 1}
The $D_{\text{KL}}^*$ defined in \eqref{eq: upper KL} serves as an upper bound for the KL divergence within $\hat{\mathcal{C}}_{\text{ML-IB}}$:
\begin{equation}\label{eq: inequality}
   D_{\text{KL}}(p_{\boldsymbol{\phi}_k}(\hat{\boldsymbol{z}}_k|\boldsymbol{x}_k) || r_k(\hat{\boldsymbol{z}}_k)) \leq  D^*_{\text{KL}}(p_{\boldsymbol{\phi}_k}(\hat{\boldsymbol{z}}_k|\boldsymbol{x}_k) || r_k(\hat{\boldsymbol{z}}_k)) 
\end{equation}
\end{lemma}
\begin{IEEEproof}
The challenge in evaluating the KL divergence term within arises from the integral involving the logarithms of the sum of Gaussians. We address this issue by employing the Log-Sum Inequality \cite{cover1999elements}, as elaborated in Appendix A.
\end{IEEEproof} 
\begin{theorem}\label{thm: 1}
For the variational upper bound  $\hat{\mathcal{C}}_{\text{ML-IB}}$ in \eqref{eq: VML-IB} and $\mathcal{C}_{\text{QML-IB}}$ in \eqref{eq: QVMDN-IB}, we establish that:
\begin{equation}\label{eq: ML-IB inequality}
    \hat{\mathcal{C}}_{\text{ML-IB}} \leq \mathcal{C}_{\text{QML-IB}}.
\end{equation}
\end{theorem}
 \begin{IEEEproof}
Referencing Lemma \ref{lemma: 1}, it is established that $D^*_{\text{KL}}$ surpasses the KL divergence component in $ \hat{\mathcal{C}}_{\text{ML-IB}}$, thereby validating the inequality in \eqref{eq: ML-IB inequality}.
 \end{IEEEproof}\par 
Following Theorem \ref{thm: 1}, we derive an approximation for the variational upper bound $\hat{\mathcal{C}}_{\text{ML-IB}}$, denoted as $\mathcal{C}_{\text{QML-IB}}$, which can be easily and quickly obtained during training. According to Monte Carlo sampling, we can obtain an unbiased estimate of the gradient for the optimization function in \eqref{eq: QVMDN-IB} and proceed with its optimization. Specifically, during the training process, given a mini-batch of data $\{(\boldsymbol{x}_{1}^{(m)}, \boldsymbol{x}_{2}^{(m)}, \dots,\boldsymbol{x}_{K}^{(m)})\}_{m=1}^M$ and sampling the channel noise $L$ times for each transmission, we acquire the following empirical estimates:
\begin{equation}\label{eq: simeq}
\begin{aligned}
        \mathcal{C}_{\text{QML-IB}} &\simeq \frac{1}{M} \sum_{m=1}^M \frac{1}{L} \sum_{l=1}^L \left\{ -\log p_{\boldsymbol{\psi}}(\boldsymbol{y}^{(m)}|\hat{\boldsymbol{z}}^{(m)}_{1:K,l}) \vphantom{\sum_{k=1}^K}\right. \\
        &\quad \left. + \sum_{k=1}^K \beta_k D_{\text{KL}}\left(p_{\boldsymbol{\phi}_k}(\hat{\boldsymbol{z}}_{k}|\boldsymbol{x}_{k}) \middle\| r_k(\hat{\boldsymbol{z}}_{k})\right) \right\},
    \end{aligned}
\end{equation}
where $\hat{\boldsymbol{z}}_{k,l}^{(m)}=\boldsymbol{z}_{k}^{(m)}+\epsilon_{k,l}^{(m)}$, and $\epsilon_{k,l}^{(m)} \sim \mathcal{N}(\boldsymbol{0}, \sigma_k^2 I)$. \par
Building on this, we devise the quantized multi-link information bottleneck (QML-IB) algorithm to generate AI models for collaborative intelligence in a multi-device scenario, which is summarized in Algorithm~\ref{al: 1}.

\begin{theorem}\label{theorem: 2}
The discrepancy between the approximation and the KL divergence is given by:
\begin{equation}
    \begin{aligned}
        &D^*_{\text{KL}}(p_{\boldsymbol{\phi}_k}(\hat{\boldsymbol{z}}_k|\boldsymbol{x}_k) || r_k(\hat{\boldsymbol{z}}_k)) - D_{\text{KL}}(p_{\boldsymbol{\phi}_k}(\hat{\boldsymbol{z}}_k|\boldsymbol{x}_k) || r_k(\hat{\boldsymbol{z}}_k)) \\
        &\leq \frac{2Td}{T+1}C+2d\delta
    \end{aligned}
\end{equation}
In this context, $T$ denotes the number of breakpoints in the quantization function and $d$ represents the dimension of the vector $\hat{\boldsymbol{z}}_k$. The constants $C$ and $\delta$ are computable and depend exclusively on the probability mass function of $\tilde{Z}_k$ given $X_k$, as well as the channel noise $\sigma_k^2$.
\end{theorem}
\begin{IEEEproof}
    By customizing the error analysis to two distinct ranges of $\hat{z}_{k,i}$, we bound the aforementioned discrepancy as outlined in Appendix B.
\end{IEEEproof}
Theorem \ref{theorem: 2} elucidates the validity of using $D_{\text{KL}}^*$ as an approximation for $D_{\text{KL}}$. It demonstrates that the discrepancy between them is under control by the values of $T$ and $d$. \par

 \begin{algorithm}[t]
 \caption{Quantized multi-link information bottleneck }  
 \label{al: 1} 
  \begin{algorithmic}[1] 
  	\renewcommand{\algorithmicrequire}{\textbf{Input:}}
	\renewcommand{\algorithmicensure}{\textbf{Output:}}
  \Require {Training dataset $\mathcal{D}$, number of iterations $P$, number of devices $K$, number of channel noise samples $L$, number of breakpoints $T$, batch size $M$, channel noise variance $\{\sigma_k^2\}_{k=1}^K$.}
  \Ensure {The optimized parameters $\{\boldsymbol{\phi}_k\}_{k=1}^K$, $\{\boldsymbol{\chi}_k\}_{k=1}^K$ and $\boldsymbol{\psi}$.}
  \For{epoch $p=1$ \textbf{to} $P$}
  \State Select a mini-batch of data $\{(\boldsymbol{x}_{1}^{(m)},\dots,\boldsymbol{x}_{K}^{(m)})\}_{m=1}^M$
  \For{$k = 1$ \textbf{to} $K$ \textbf{and} $m = 1$ \textbf{to} $M$}
  \State Compute the feature vector $\boldsymbol{z}_{k}^{(m)}$ based on \eqref{eq: conditional x_k to z_k}
  \State Compute the quantized vector $\tilde{\boldsymbol{z}}_{k}^{(m)}$ based on \eqref{eq: soft-quantization}
  \State Sample the noise $\{\epsilon_{k,l}^{(m)}\}_{l=1}^L \sim \mathcal{N}(\boldsymbol{0}, \sigma_k^2 I)$
  \State Compute the received vector based on  \Statex \qquad \quad 
$\hat{\boldsymbol{z}}_{k,l}^{(m)}=\tilde{\boldsymbol{z}}_{k}^{(m)}+\epsilon_{k,l}^{(m)}$
  \EndFor
  \State Compute the KL-divergence based on \eqref{eq: upper KL}
  \State Compute the loss $\mathcal{C}_{\text{QML-IB}}$ based on \eqref{eq: simeq}
  \State Update parameters through backpropagation 
  \EndFor
\end{algorithmic}  
\end{algorithm}

\section{Numerical Results and Discussions}
This section evaluates the performance of our proposed algorithm in device-network collaborative intelligence. 
\subsection{Experimental Settings}
We selected the MNIST\cite{lecun1998gradient} and CIFAR-10\cite{krizhevsky2009learning} as benchmark datasets for classification tasks. We conducted numerical experiments for both of our proposed QML-IB algorithm and a benchmark. With limited research in multi-link collaborative intelligence, we select the method VDDIB-SR\cite{shao2022task} as our baseline for comparison. Our algorithm and the baseline both feature a quantization framework. Note that to ensure a comprehensive experimental comparison, we simulate their non-quantized versions as well. 
\par
Following the method in \cite{whang2021neural}, we split each image into $K$ equal, non-overlapping sub-images, with each device assigned a sub-image. The hyperparameters \(\beta_k\) (for \(k=1,\dots,K\)) in \eqref{eq: ML-IB} are set within the range of \([10^{-5}, 10^{-1}]\), and the grid search is utilized to determine their optimal values for the MNIST and CIFAR-10 datasets \cite{xie2023robust}. 
Following \cite{ye2021autoencoder}, we set the constant $\gamma$ in \eqref{eq: soft-quantization} to 10. For the constant $T$, set at 15, further discussion is available in Appendix \ref{Appendx: D.A}. For consistency, we followed the setup from \cite{shao2021learning} with a 12.5kHz bandwidth and 9,600 baud symbol rate. We adopt the same neural network
backbone for all algorithms, as depicted in Tables \ref{tab:DNN_MNIST} and \ref{tab:DNN_CIFAR}. The Peak Signal-to-Noise Ratio (PSNR) for channel \( k \) is:
\begin{equation}
\text{PSNR} = 10 \log \left( \frac{P_{\text{max}}}{\sigma_k^2} \right) \text{dB},
\end{equation}
where \( P_{\text{max}}, \) the signal's maximum power, is set to 1 in our experiments, as each component of the encoded vector \( z_k \) is limited to \( |z_{k,i}| \leq 1\).

\begin{table}[t]
    \caption{THE DNN ARCHITECTURE FOR MNIST CLASSIFICATION}
    \centering
    \begin{tabular}{c|c|c}
    \hline
    &  \textbf{Layer} &  \textbf{Outputs} \\
    \hline
  \textbf{Device-side Network}& FC Layer + Tanh & d\\ \hline
   
    \multirow{3}{*}{\textbf{Server-side Network}}  & FC Layer + ReLU & 1024 \\ 
    &  FC Layer + ReLU & 256 \\ 
    & FC Layer + Softmax & 10 \\
    \hline
    \end{tabular}
    \label{tab:DNN_MNIST}
\end{table}

\begin{table}[t]
\caption{THE DNN ARCHITECTURE FOR CIFAR-10 CLASSIFICATION.}
\centering
\scriptsize{
\begin{tabular}{c|c|c}
\hline
\textbf{Network} & \textbf{Layer} & \textbf{Outputs} \\
\hline
\multirow{3}{*}{\textbf{Device-side Network}} & Conv Layer  $\times$ 2 + ResNet Block& $128 \times 32 \times 32$ \\
& Conv Layer  $\times$ 2 + ResNet Block& $512 \times 4 \times 4$ \\
& Conv Layer  & $4 \times 4 \times 4$ \\
& Reshape + FC Layer + Tanh & $d$ \\
\hline
\multirow{4}{*}{\textbf{Server-side Network}} & FC Layer + Reshape & $4\times 4 \times 4$ \\
& Conv Layer$\times$ 2  & $512 \times 4 \times 4$ \\
& ResNet Block $\times$ 2  & $512 \times 4 \times 4$ \\
& Pooling Layer & $512$ \\
& FC Layer + Softmax & $10$ \\
\hline
\end{tabular}}
\label{tab:DNN_CIFAR}
\end{table}

\subsection{Experimental Results}
\begin{figure}[t]
\raggedright
\includegraphics[width=\linewidth]{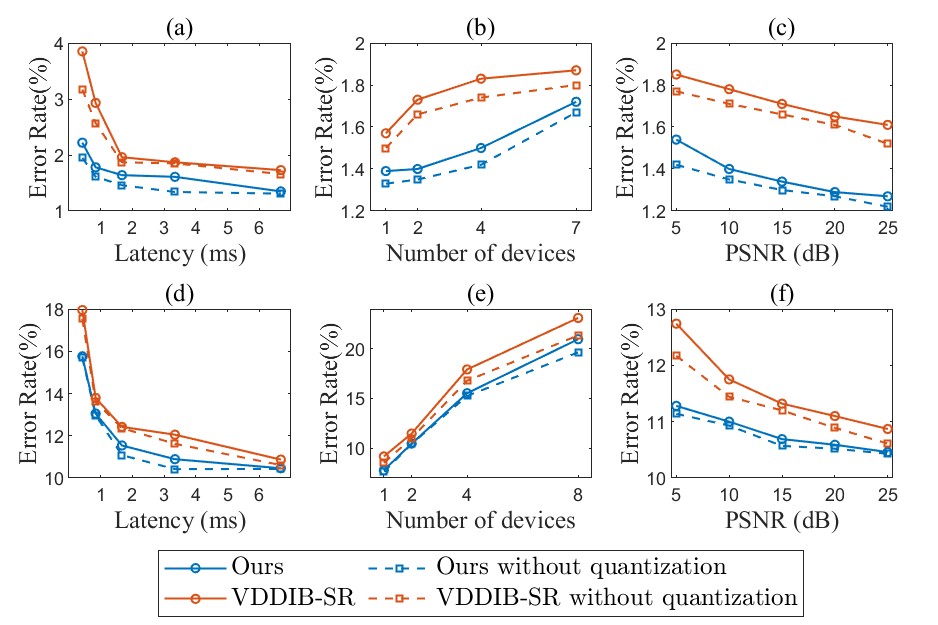}
\caption{Performance comparison between our algorithm and VDDIB-SR in different experimental settings. Evaluations on MNIST Dataset in Figs.~\ref{fig: experimental results}(a), (b), (c) and on CIFAR-10 Dataset in Figs.~\ref{fig: experimental results}(d), (e), (f).}
\label{fig: experimental results}
\end{figure}
In our experiments, we evaluate the error rate of our algorithm and the benchmark VDDIB-SR for processing AI tasks, in terms of three key aspects: communication latency, number of devices, and PSNR. First, we standardize the PSNR at 10 dB in a two-device configuration to evaluate the algorithm performance under varying communication latencies, detailed in Figs.~\ref{fig: experimental results}(a) and (d). We then evaluate how well the algorithms adapt to different numbers of devices while keeping the communication latency below 6 ms, as shown in Figs.~\ref{fig: experimental results}(b) and (e). Finally, to test the robustness of the algorithms, we vary the PSNR while maintaining the device count at 2, observing their performance across different channel noise in Figs.~\ref{fig: experimental results}(c) and (f).\par
Results from Fig.~\ref{fig: experimental results} show that our algorithm consistently exhibits a lower error rate than VDDIB-SR in various experimental scenarios, indicating superior performance in multi-device collaborative settings. This confirms the superiority of the ML-IB method and the well-designed performance metric $\mathcal{C}_{\text{ML-IB}}$. Specifically, Figs.~\ref{fig: experimental results}(a) and (d) show that our algorithm achieves lower error rates with reduced latency. Figs.~\ref{fig: experimental results}(b) and (e) demonstrate its ability to effectively coordinate an increasing number of devices. In addition, Figs.~\ref{fig: experimental results}(c) and (f) highlight its consistent outperformance under varying PSNR, validating its robustness to different channel conditions.
\par
Furthermore, experimental results indicate that our algorithm maintains an error rate close to its unquantized version. This reflects that our quantization, tailored to real-world scenarios and compatible with digital communication systems, incurs only negligible performance loss. Notably, our algorithm, even with the quantization scheme applied, still outperforms the baseline's non-quantized version. This underscores the ability of our QML-IB algorithm to provide strong performance assurance and confirms its significant practical value.

\section{Conclusion}
In this work, we investigate the establishment of collaborative intelligence across multiple wireless links. We introduce an
ML-IB scheme with a novel performance metric $\mathcal{C}_{\text{ML-IB}}$.
To fulfill practical communication requirements, we design a quantizer that offers adaptable management of quantization parameters, including bit depth, breakpoint, and amplitude. 
Using variational methods and the Log-Sum Inequality, we derive an approximation for $\mathcal{C}_{\text{ML-IB}}$ to make the ML-IB scheme computable. We also provide an error analysis to demonstrate the validity of this approximation. On this basis, we propose the QML-IB algorithm for AI model generation. Numerical experiments validate the effectiveness of our algorithm and the superiority of the performance metric. We want to stress that our proposed approach may be applicable to various communication channels, not just the additive white Gaussian noise channel that we discussed in this paper. For instance, it could be used on the Rayleigh Fading Channel and the Rician Fading Channel \cite{goldsmith2005wireless}. This is an interesting topic that warrants further research.

 \IEEEtriggeratref{16}


\clearpage
\bibliographystyle{IEEEtran}
\bibliography{bibliofile}

\clearpage
\appendices

\section{Proof of Lemma \ref{lemma: 1}}
Considering the conditional independence of $\hat{z}_{k,i}$ and the isotropy of the variational marginal distribution, we have: 
\begin{equation*}
    \begin{aligned}
    D_{\text{KL}}(p_{\boldsymbol{\phi}_k}(\hat{\boldsymbol{z}}_k|\boldsymbol{x}_k)||r_k(\hat{\boldsymbol{z}}_k))  = 
    \int p_{\boldsymbol{\phi}_k}(\hat{\boldsymbol{z}}_k|\boldsymbol{x}_k)\log\frac{p_{\boldsymbol{\phi}_k}(\hat{\boldsymbol{z}}_k|\boldsymbol{x}_k)}{r(\hat{\boldsymbol{z}}_k)} d\hat{\boldsymbol{z}}_k&\\
     = \sum_{i=1}^d \int p_{\boldsymbol{\phi}_k}(\hat{z}_{k,i}|\boldsymbol{x}_k)\log \frac{p_{\boldsymbol{\phi}_k}(\hat{z}_{k,i}|\boldsymbol{x}_k)}{r_k(\hat{z}_{k,i})}d\hat{z}_{k,i}&.\\
    \end{aligned}
\end{equation*}\par 
Here, $p_{\boldsymbol{\phi}_k}(\hat{z}_{k,i}|\boldsymbol{x}_{k})=\sum_{t=1}^{T+1} P_{k,i}(c_{k,t})f_{\epsilon}(\hat{z}_{k,i}-c_{k,t})$ and $r_k(\hat{\boldsymbol{z}}_k) = \mathcal{N}(\hat{\boldsymbol{z}}_k|\boldsymbol{0},\boldsymbol{I})$. Then we define $a_{i,t} = P_{k,i}(c_{k,t})f_{\epsilon}(\hat{z}_{k,i}-c_{k,t})$ and $b_{i,t} = \frac{1}{T+1}r_k(\hat{\boldsymbol{z}}_{k})$. Applying the Log-Sum Inequality \cite{cover1999elements}, we have:
\begin{equation}
\left(\sum_{t=1}^{T+1} a_{i,t}\right) \log \frac{\sum_{t=1}^{T+1}a_{i,t}}{\sum_{t=1}^{T+1} b_{i,t}} \leq
\sum_{t=1}^{T+1} a_t \log \frac{a_{i,t}}{b_{i,t}}.
\end{equation}
Consequently, this leads to the inequality:
\begin{equation*}
\begin{aligned}
&\sum_{t=1}^{T+1} P_{k,i}(c_{k,t})f_{\epsilon}(\hat{z}_{k,i}-c_{k,t}) \log \frac{\sum_{t=1}^{T+1} P_{k,i}(c_{k,t})f_{\epsilon}(\hat{z}_{k,i}-c_{k,t})}{r(\hat{z}_{k,i})} \\
&\leq \sum_{t=1}^{T+1} P_{k,i}(c_{k,t})f_{\epsilon}(\hat{z}_{k,i}-c_{k,t}) \log \frac{P_{k,i}(c_{k,t})f_{\epsilon}(\hat{z}_{k,i}-c_{k,t})}{\frac{1}{T+1}r(\hat{z}_{k,i})}.\\
\end{aligned}    
\end{equation*}
Applying the formula for calculating the cross-entropy between two Gaussian distributions and integrating the right-hand side of the inequality, we derive
\begin{equation}\label{eq: appendix3}
    \begin{aligned}
    &\int p_{\boldsymbol{\phi}_k}(\hat{z}_{k,i}|\boldsymbol{x}_k)\log \frac{p_{\boldsymbol{\phi}_k}(\hat{z}_{k,i}|\boldsymbol{x}_k)}{r_k(\hat{z}_{k,i})}d\hat{z}_{k,i} \\ 
    &\leq -H(\tilde{Z}_{k,i})-\frac{1}{2}\log{\sigma_{k}^2}+\frac{1}{2}\sigma_{k}^2-\frac{1}{2}+\sum_{t=1}^{T+1} p_{k,i}(c_t)(c_t)^2,
    \end{aligned}
\end{equation}
Next, we sum over both sides of inequality \eqref{eq: appendix3} for $i=1$ to $d$, leading to inequality \eqref{eq: inequality} and completing the proof.\par

\section{Proof of Theorem \ref{theorem: 2}}
Let \(C_{i,t} = (T+1)P_{k,i}(c_{k,t}) \frac{1}{\sigma_k} e^{\frac{- c_{k,t}^2}{2(1-\sigma_k^2)}}\) for \(i = 1, \ldots, d\) and \(t = 1, \ldots, T+1\), with \(C_{\text{max}} = \max_{i,t} \{C_{i,t}\}\) and \(C_{\text{min}} = \min_{i,t} \{C_{i,t}\}\). Define \(C_0 = \max (||\log C_{\text{min}}||, ||\log C_{\text{max}}||)\), \(\varepsilon = \frac{(1+3\sigma_k+\frac{1}{1-\sigma_k^2})^2}{\frac{2\sigma_k^2}{1-\sigma_k^2}}\) and $\delta =\max (\frac{1}{e}, C_{\text{max}}e^{-\varepsilon}\log(C_{\text{max}}e^{-\varepsilon}))$.
We continue using the notation from Theorem \ref{thm: 1}, defining $f(x)=x\log x$, $\alpha_{i,t} = \frac{b_{i,t}}{\sum_{t=1}^{T+1}b_{i,t}}$, and $\beta_{i,t}=\frac{a_{i,t}}{b_{i,t}}$. Thus,
\begin{equation*}
\begin{aligned}
&D^*_{\text{KL}}(p_{\boldsymbol{\phi}_k}(\hat{\boldsymbol{z}}_k|\boldsymbol{x}_k) || r_k(\hat{\boldsymbol{z}}_k)) - D_{\text{KL}}(p_{\boldsymbol{\phi}_k}(\hat{\boldsymbol{z}}_k|\boldsymbol{x}_k) || r_k(\hat{\boldsymbol{z}}_k)) \\
&=\sum_{i=1}^d \int (\sum_{t=1}^{T+1}{b_{i,t}}) (\sum_{t=1}^{T+1} {\alpha_{i,t}}f(\beta_{i,t}) - f(\sum_{t=1}^{T+1} \alpha_{i,t} \beta_{i,t})) d\hat{z}_{k,i}.
\end{aligned}
\end{equation*}
Given the expression of $\beta_{i,t}$:
\begin{equation*}
\begin{aligned}
    \beta_{i,t} = \frac{a_{i,t}}{b_{i,t}} = (T+1)P_{k,i}(c_{k,t}) \frac{1}{\sigma_k} e^{\frac{- c_{k,t}^2}{2(1-\sigma_k^2)}} e^{-\frac{(\hat{z}_{k,i}-\frac{c_{k,t}}{1-\sigma_k^2})^2}{\frac{2\sigma_k^2}{1-\sigma_k^2}}},
\end{aligned}
\end{equation*}
when $\hat{z}_{k,i} \in (-1-3\sigma_k,1+3\sigma_k)$, the bounds for $\beta_{i,t}$ can be described as:
\begin{equation}\label{eq: bound for beta_t}
C_{i,t} e^{-\varepsilon} \leq \beta_{i,t} \leq C_{i,t}.
\end{equation}
Furthermore,
\begin{equation*}
    \begin{aligned}
        &|| (\sum_{t=1}^{T+1}{b_{i,t}}) (\sum_{t=1}^{T+1} {\alpha_{i,t}}f(\beta_{i,t}) - f(\sum_{t=1}^{T+1} \alpha_{i,t} \beta_{i,t}))|| \\
        &\leq  || \sum_{t=1}^{T+1}{b_{i,t}}|| \cdot 
        \sum_{t=1}^{T+1}\left(||\alpha_{i,t}||\cdot 
        ||f(\beta_{i,t})-f(\sum_{t=1}^{T+1} \alpha_{i,t} \beta_{i,t})||\right).
    \end{aligned}
\end{equation*}
Applying the Lagrange median theorem, we find
\begin{equation*}
\begin{aligned}
&||f(\beta_{i,t})-f(\sum_{t=1}^{T+1} \alpha_{i,t} \beta_{i,t})|| 
=||f^{'} (\xi)|| \cdot || \beta_{i,t} - \sum_{t=1}^{T+1} \alpha_{i,t} \beta_{i,t})  ||\\
&\leq  ||f^{'} (\xi)|| \sum_{j=1}^{T+1} \left(||\alpha_{i,j}|| \cdot ||\beta_{i,t} - \beta_{i,j}|| \right) \\
&\leq  \frac{1}{T+1}||f^{'} (\xi)||  (\sum_{j\neq t}||\beta_{i,j}||+T||\beta_{i,t}||),
\end{aligned}
\end{equation*}
where $\xi \in [\beta_{i,t}, \sum_{t=1}^{T+1} \alpha_{i,t} \beta_{i,t})]$ or $\xi \in [\sum_{t=1}^{T+1} \alpha_{i,t} \beta_{i,t}, \beta_{i,t})]$. 
Given $f(x) = x\log x$ with the derivative  $f^{'}(x)=\log x+1$ and considering the range of $\beta_{i,t}$ as defined in \eqref{eq: bound for beta_t}, we can deduce that 
\begin{equation*}
\begin{aligned}
||f^{'} (\xi)|| < C_0 + \log \varepsilon +1.
\end{aligned}
\end{equation*}
Consequently,
\begin{equation*}
\begin{aligned}
    &|| (\sum_{t=1}^{T+1}{b_{i,t}}) (\sum_{t=1}^{T+1} {\alpha_{i,t}}f(\beta_{i.t}) - f(\sum_{t=1}^{T+1} \alpha_{i,t} \beta_{i,t}))|| \\
    & \leq
    \frac{2T}{T+1}(C_0+\log \varepsilon +1)(\sum_{t=1}^T a_{i,t}).
\end{aligned}
\end{equation*}
Then 
\begin{equation}\label{eq: appendix1}
\begin{aligned}
&\int_{-1-3\sigma_k}^{1+3\sigma_k} (\sum_{t=1}^{T+1}{b_{i,t}}) (\sum_{t=1}^{T+1} {\alpha_{i,t}}f(\beta_{i,t}) - f(\sum_{t=1}^{T+1} \alpha_{i,t} \beta_{i,t})) d\hat{z}_{k,i}\\
&\leq \frac{2T}{T+1}(C_0+\log \varepsilon +1).
\end{aligned}
\end{equation}
For $\hat{z}_{k,i}$ within the ranges $(-\infty,-1-3\sigma_k)$ or $(1+3\sigma_k,\infty)$, it is deduced that $0< \beta_{i,t} \leq C_{i,t}e^{-\varepsilon}$. This implies $|f(\beta_{i,t})|$ is less than the maximum of ${\frac{1}{e},|C_{i,t}e^{-\varepsilon} \log C_{i,t}e^{-\varepsilon}|}$, which in turn is bounded above by $\delta$, applicable for $i=1,\dots, d$ and $t=1,\dots,T+1$.
So
\begin{equation}\label{eq: appendix2}
\begin{aligned}
&\left(\int_{-\infty}^{-1-3\sigma_k}+\int_{1+3\sigma_k}^{\infty}\right) \\
&\quad (\sum_{t=1}^{T+1}{b_{i,t}}) \left(\sum_{t=1}^{T+1} {\alpha_{i,t}}f(\beta_{i,t}) - f\left(\sum_{t=1}^{T+1} \alpha_{i,t} \beta_{i,t}\right)\right) d\hat{z}_{k,i}\\
&\leq 2\delta.
\end{aligned}
\end{equation}
Therefore, based on the inequalities \eqref{eq: appendix1} and \eqref{eq: appendix2}, we can deduce that:
\begin{equation}
\begin{aligned}
&\int (\sum_{t=1}^{T+1}{b_{i,t}}) (\sum_{t=1}^{T+1} {\alpha_{i,t}}f(\beta_{i,t}) - f(\sum_{t=1}^{T+1} \alpha_{i,t} \beta_{i,t})) d\hat{z}_{k,i}\\ 
&\leq \frac{2T}{T+1}(C_0 + \log \varepsilon + 1)+2d\delta.
\end{aligned}
\end{equation}
We denote $C = C_0 + \log \varepsilon + 1$, then the actual error can be expressed as:
\begin{equation}
    \begin{aligned}
        &D^*_{\text{KL}}(p_{\boldsymbol{\phi}_k}(\hat{\boldsymbol{z}}_k|\boldsymbol{x}_k) || r_k(\hat{\boldsymbol{z}}_k)) - D_{\text{KL}}(p_{\boldsymbol{\phi}_k}(\hat{\boldsymbol{z}}_k|\boldsymbol{x}_k) || r_k(\hat{\boldsymbol{z}}_k)) \\
        &\leq \frac{2Td}{T+1}C+2d\delta.
    \end{aligned}
\end{equation}

\begin{figure}[t]
\centering
\begin{subfigure}{.24\textwidth}
  \centering
  \includegraphics[width=\linewidth]{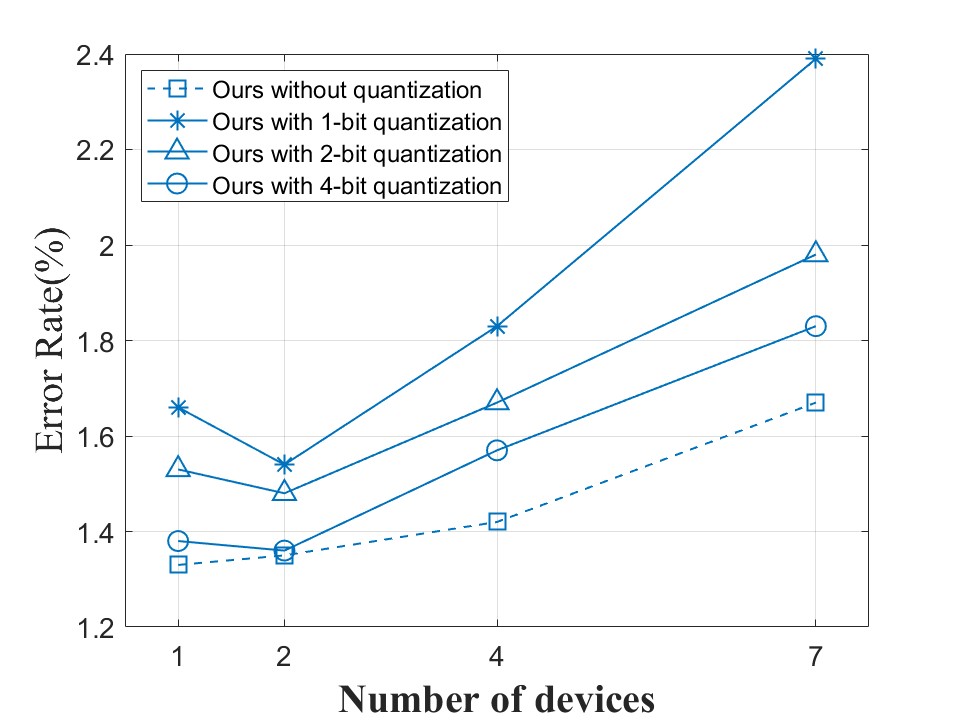}
  \caption{MNIST dataset.}
  \label{fig:MNSIT_ablation_new_sub1}
\end{subfigure}%
\begin{subfigure}{.24\textwidth}
  \centering
  \includegraphics[width=\linewidth]{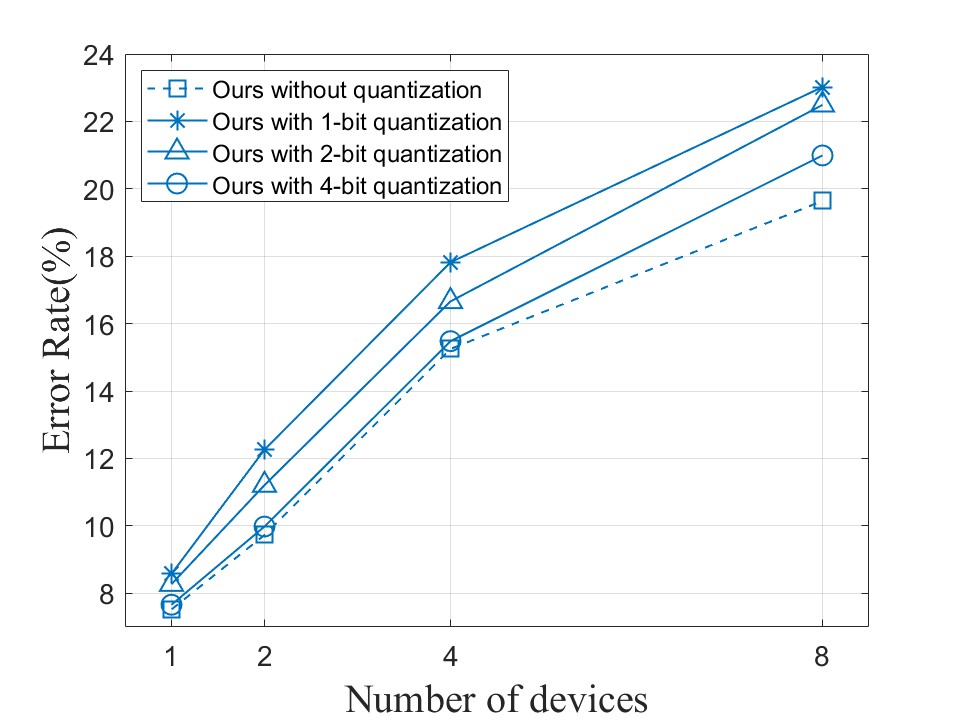}
  \caption{CIFAR-10 dataset.}
  \label{fig:CIFAR_ablation_new_sub1}
\end{subfigure}
\caption{Impact of the quantization bit depth on performance}
\label{fig: quantization_ablation}
\end{figure}

\begin{figure}[t]
\centering
\begin{subfigure}{.24\textwidth}
  \centering
  \includegraphics[width=\linewidth]{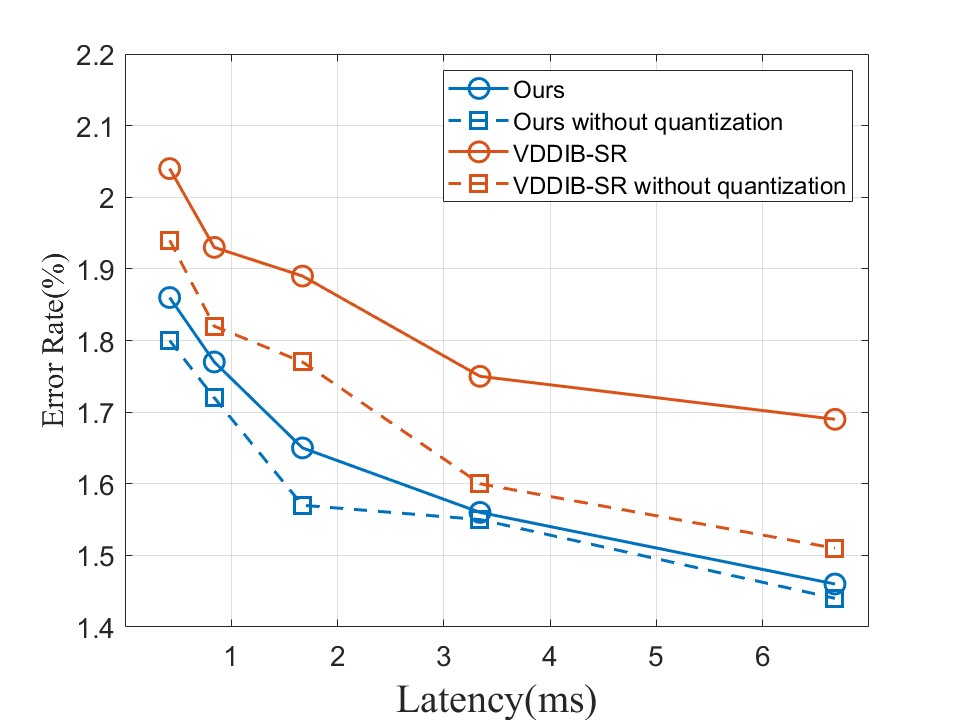}
  \caption{MNIST dataset.}
\end{subfigure}%
\begin{subfigure}{.24\textwidth}
  \centering
  \includegraphics[width=\linewidth]{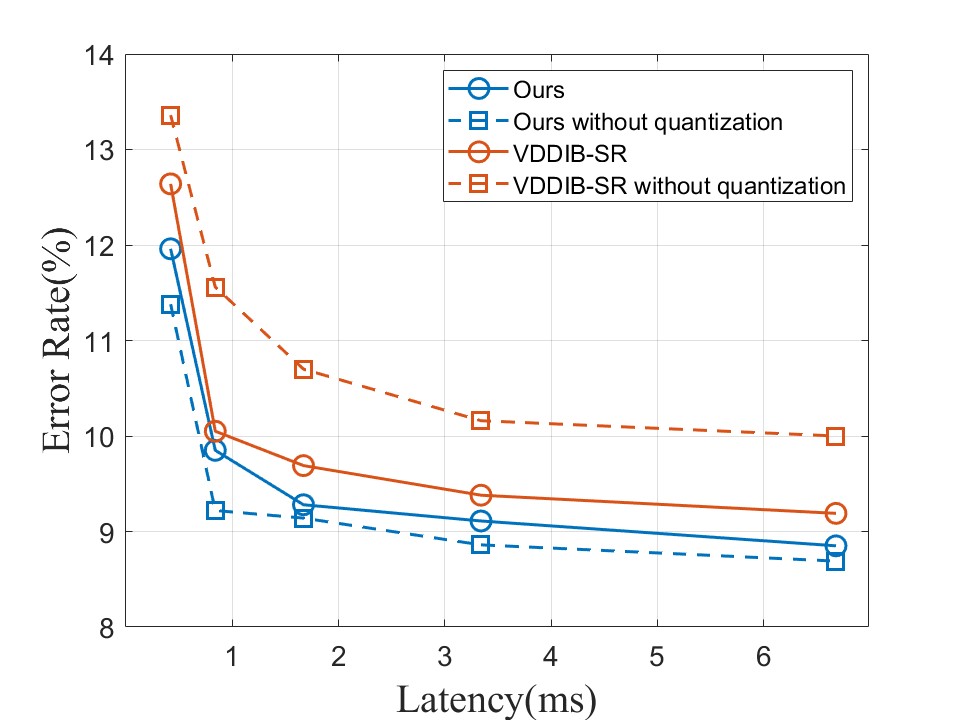}
  \caption{CIFAR-10 dataset.}
\end{subfigure}
\caption{Performance in partial overlap of original input data}
\label{fig: overlab}
\end{figure}

\section{Ablation Study}\label{Appendx: D.A}

In this subsection, we investigate the impact of our quantization framework on inference performance with different numbers of quantization bits. Furthermore, we further analyze the performance of our algorithm when the raw data received by different devices partially overlap.

\subsection{Number of Quantization Bits}
We set the PSNR to 10 dB and the communication latency remains below 6 ms. Within our proposed quantization framework, we compare four versions: quantization with 2, 4, and 16 discrete values, corresponding to 1, 2, and 4 quantization bits, respectively, and a version without quantization. As illustrated in Fig. \ref{fig: quantization_ablation}, it is evident that the error rate decreases as the number of discrete quantization values increases, progressively converging towards the performance of the non-quantized version. Remarkably, with only a 4-bit quantization of our framework, the error is already very close to the version without quantization. This clearly demonstrates that the performance degradation introduced by our quantization is negligible.

\subsection{Scenario with Overlapping Raw Input Data}
We divide each image into $K$ equal sub-images, ensuring a 50\% overlap between adjacent sub-images, with each one allocated to a distinct device, as detailed in Fig.~\ref{fig: overlab}. When there is partial overlap in the input data, the error rate generated by our algorithm is consistently lower than the baseline, further demonstrating the applicability of our algorithm in diverse collaborative intelligence scenarios. Additionally, we find that in scenarios where data overlap occurs, the error rate of our algorithm is lower than in non-overlapping cases. This indicates that the intersection of original input data across different devices can better facilitate the extraction of task-relevant information, consequently enhancing the performance of AI tasks.

\end{document}